\documentclass[pra,twocolumn,aps,amssymb,footinbib]{revtex4}

\usepackage{amsfonts}
\usepackage{amsmath}
\usepackage{amssymb}
\usepackage{graphicx}%
\providecommand{\U}[1]{\protect\rule{.1in}{.1in}}

\begin{document}
\title{Scalable Architecture for a Room Temperature Solid-State Quantum Information Processor}
\date{\today}
\author{N. Y. Yao$^{1\dagger*}$, L. Jiang$^{2\dagger}$, A. V. Gorshkov$^{1,2\dagger}$, P. C. Maurer$^{1}$, G. Giedke$^{3}$, J. I. Cirac$^{3}$, M. D. Lukin$^{1}$}

\affiliation{$^{1}$Physics Department, Harvard University, Cambridge, MA 02138}
\affiliation{$^{2}$Institute for Quantum Information, California Institute of Technology, Pasadena, CA 91125}
\affiliation{$^{3}$Max-Planck-Institut f\"ur Quantenoptik, Hans-Kopfermann-Strasse 1, Garching, D-85748, Germany}
\affiliation{$^{\dagger}$These authors contributed equally to this work}
\affiliation{$^{*}$e-mail: nyao@fas.harvard.edu}

\begin{abstract}
The realization of a scalable quantum information processor has emerged over the past decade as one of the central challenges at the interface of fundamental science and engineering. Much progress has been made towards this goal. Indeed, quantum operations have been demonstrated on several trapped ion qubits, and other solid-state systems are approaching similar levels of control. Extending these techniques to achieve fault-tolerant operations in larger systems with more qubits remains an extremely challenging goal, in part, due to the substantial technical complexity of current implementations. Here, we propose and analyze an architecture for a scalable, solid-state quantum information processor capable of operating at or near room temperature.  The architecture is applicable to realistic conditions, which include disorder and relevant decoherence mechanisms, and includes a hierarchy of control at successive length scales. Our approach is based upon recent experimental advances involving Nitrogen-Vacancy color centers in diamond and will provide fundamental insights into the physics of non-equilibrium many-body quantum systems. Additionally, the proposed architecture may greatly alleviate the stringent constraints, currently limiting the realization of scalable quantum processors. 
\end{abstract}
\maketitle

\vspace{45mm}

Nitrogen-Vacancy (NV) color centers in diamond stand out among other promising qubit implementations~\cite{Loss98, Raussendorf01, Duan04, Benjamin06,JTL07, Kane98, Makhlin01} in that their electronic spins can be individually polarized, manipulated and optically detected under room-temperature conditions.  Recent advances involving the quantum manipulation of such crystal defects have allowed researchers to achieve sub-diffraction limited resolution and dipole-coupling mediated entanglement between neighboring NV electronic spins~\cite{Epstein05, Neumann10b, Childress06, Balasubramanian09, Weber10, Fuchs10, Balasubramanian08, Rittweger09, JDL08, JHML09, Neumann10, Togan10}. Despite such substantial developments in this and other experimental systems, it remains unclear whether these pieces can be combined into a scalable quantum information processor (QIP) capable of operating under ambient, room temperature conditions. 

In what follows, we describe and analyze a feasible architecture for a diamond-based quantum information processor.
Our approach makes use of an array of single NV centers, created through implantation of  ions and subsequent annealing~\cite{Balasubramanian09, Tallairea06}. Each NV center constitutes an individual quantum register containing a nuclear spin and a localized electronic spin. The nuclear spin, which has a long coherence time, serves as the memory qubit, storing quantum information, while the electronic spin will be used to initialize, read out, and mediate coupling between nuclear spins of adjacent registers. Magnetic dipole interactions allow for coherent coupling between NV centers spatially separated by tens of nanometers. While in principle, a perfect array of NV centers would enable scalable quantum information processing, in practice, the finite creation efficiency of such centers, along with the requirements for parallelism, necessitate the coupling of registers separated by significantly larger distances. To overcome this challenge, we show that the coupling between NV centers can be mediated by an optically un-addressable ``dark'' spin chain data bus (DSCB). For concreteness, within our architecture, we will consider the specific implementation of such a DSCB  by utilizing implanted Nitrogen impurities (P3 centers) with spin 1/2, as shown in Fig.~1a.~\cite{Epstein05, Reynhardt98}.


\begin{figure}
\centering
\includegraphics[width=3.4in]{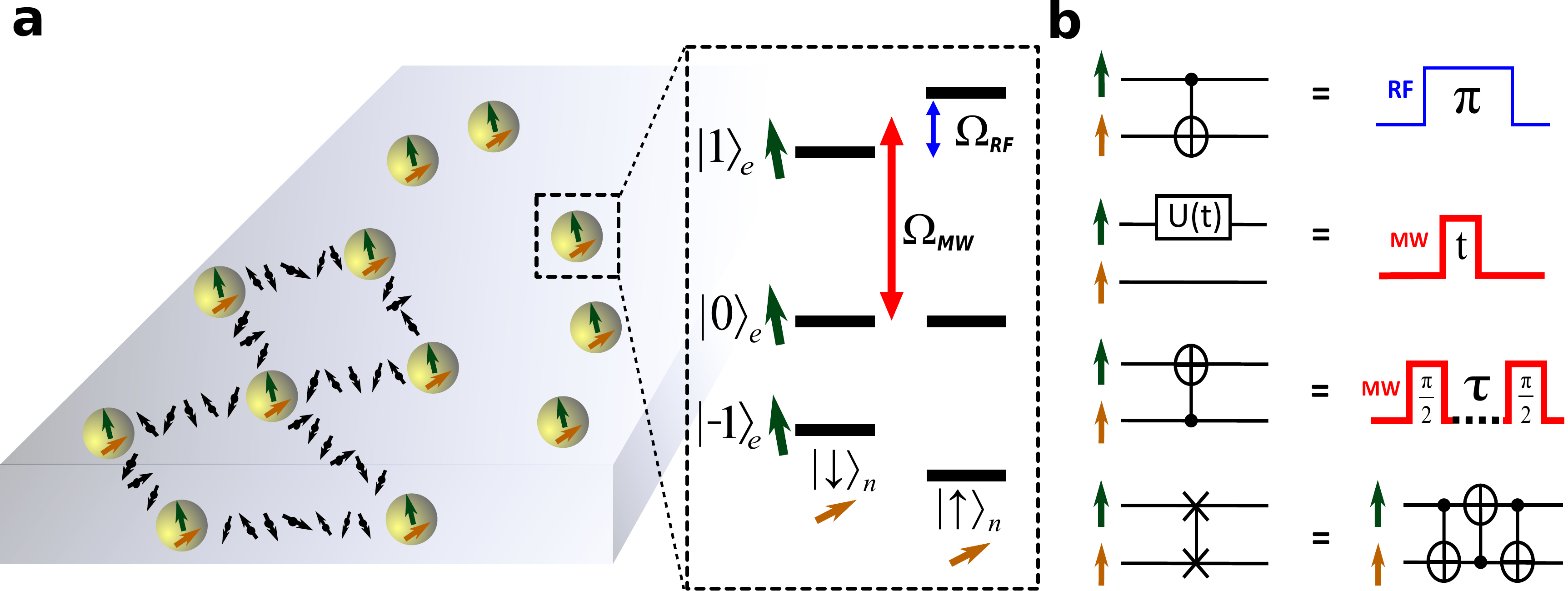}
\caption{Schematic representation of individual NV registers within bulk diamond.~(a) Each NV register contains a nuclear spin $I=1/2$ (yellow), providing quantum memory, and an electronic spin $S=1$ (green). Dark spins (black) represent elements of an optically un-addressable spin chain which coherently couples spatially separated NV registers. The NV level structure (in a high $B$ field) is shown. A resonant microwave ($\Omega_{MW}$) pulse coherently transfers the electronic spin of the register from $\left\vert0\right\rangle _{e}$ to $\left\vert1\right\rangle _{e}$; subsequent manipulation of the nuclear spin is accomplished through an RF pulse ($\Omega_{RF}$). The far detuned  $\left\vert-1\right\rangle _{e}$ state can be neglected to create an effective two-qubit register. However, the full three level NV structure will be utilized in horizontal DSCB mediated coherent coupling of NV registers. (b) A universal set of two-qubit gates can easily be achieved with only MW and RF controls~\cite{Cappellaro09}. Electronic spin manipulation can be accomplished with a MW field, where $t$ represents the duration of the MW pulse. By exploiting the hyperfine coupling between the electronic and nuclear spin, one can achieve controlled-NOT operations conditioned on either spin. In particular, a C$_{e}$NOT$_{n}$ gate can be accomplished by utilizing a RF $\pi$-pulse, which flips the nuclear spin conditioned on the electronic spin being in $|1 \rangle_{e}$.  Similarly, a C$_{n}$NOT$_{e}$ gate can be accomplished by utilizing the hyperfine interaction to generate a controlled-phase (CP) gate, where $\tau$ represents the duration of the wait time required to achieve such a hyperfine-driven CP gate. Performed between two single-qubit Hadamard gates ($\pi/2$-pulses) on the electronic spin, such a CP gate generates the desired C$_{n}$NOT$_{e}$ gate. Finally, combining the C$_{e}$NOT$_{n}$ and C$_{n}$NOT$_{e}$ gates allows for the execution of a SWAP gate.} 
\end{figure}

\section*{The NV\ Qubit Register}

Single NV registers contain a spin triplet electronic ground state ($S=1$) and can be optically pumped and initialized to the $\left\vert0\right\rangle _{e}$ spin state, which has no magnetic dipole coupling with other NV registers or impurities. After optical initialization, the electronic spin of each register remains in the $\left\vert0\right\rangle _{e}$ state, unless coherently transferred to the $\left\vert 1 \right\rangle _{e}$ state by a resonant microwave (MW) pulse, as depicted in Fig.~1a~\cite{Childress06, Balasubramanian09, Weber10, Fuchs10}. The nuclear spin associated with Nitrogen atoms ($I=1/2$ for ${}^{15}$N) possesses an extremely long coherence time and will serve as the memory qubit in our system~\cite{Dutt07}; manipulation of the nuclear spin is accomplished with RF pulses~\cite{Cappellaro09}.  The Hamiltonian governing the electronic and nuclear spin of the NV register is
\begin{equation}
H_{e,n}=\Delta_{0}S_{z}^{2}+\mu_{e}B S_{z}+\mu_{n}BI_{z}+AS_{z}I_{z},
\end{equation}
\noindent with zero-field splitting $\Delta_{0}=2.87$GHz, electronic spin gyromagnetic ratio $\mu_{e}=-2.8$MHz/Gauss, nuclear spin gyromagnetic ratio $\mu_{n}=-0.43$ kHz/Gauss, and hyperfine coupling $A=3.0$ MHz~\cite{Childress06}. The application of a magnetic field along the NV-axis ($\hat{z}$) ensures full addressability of the two-qubit system, resulting in the energy levels shown in Fig.~1a. A universal set of two qubit quantum operations can easily be achieved with only MW and RF controls, as shown in Fig.~1b~\cite{Cappellaro09}. 

Furthermore, it is possible to selectively readout the state of the NV register; for example, to readout the nuclear qubit of a register, we apply a C$_{n}$NOT$_{e}$ gate to couple the electronic and nuclear spins, thereby allowing for readout of the electronic spin based on fluorescence detection. In the case where NV registers are separated by sub-optical-wavelength distances, the readout of registers will be complicated by the strong fluorescence background from neighboring NV centers. To suppress this background fluorescence, a red donut beam can be used, with its minimum located at the particular NV center being read out~\cite{Rittweger09}. Thus, while the fluorescence signal from the NV register located at the minimum persists, the remaining illuminated registers will be dominated by the stimulated emission induced by the red donut beam. In addition to suppressing the background noise, the red donut beam can also suppress the nuclear decoherence of the remaining NV registers, by reducing the amount of time these registers spend in the excited electronic state. After each round of fluorescence detection, the electronic spin is polarized to the $\left\vert 0\right\rangle _{e}$ state, while the $I_{z}$ component of the nuclear spin, a quantum non-demolition observable, remains unchanged~\cite{Haroche06book}. Therefore, it is possible to repeat this readout procedure multiple times in order to improve the readout fidelity~\cite{JHML09, Neumann10}. A strong magnetic field  $B_{z,0}\sim 1$ Tesla along the NV axis should be used to decouple the electronic and nuclear spins in order to achieve high fidelity single shot readout of NV registers~\cite{Neumann10}. In addition to sub-wavelength readout, optical donut beams also introduce the possibility of selectively manipulating individual NV registers with subwavelength resolution. While un-illuminated NV centers may respond to a resonant MW pulse, illuminated registers undergo a strong optical cycling transition which suppresses their response to microwave pulses due to the quantum Zeno effect~\cite{Misra77, Itano90, Maurer09}. 

\section*{Approach to Scalable Architecture}

One of the key requirements for fault-tolerant quantum computation is the ability to perform parallel gate operations. In our approach, this is achieved  by considering a hierarchy of controllability. The lowest level of the hierarchy consists of an individual optically addressable \emph{plaquette} with horizontal and vertical spatial dimensions $\sim 100-500$nm, containing a single computational NV register, as shown in Fig. 2a. The plaquette dimensions are chosen such that register control and readout can be achieved using conventional far-field or sub-wavelength optical techniques~\cite{Childress06, Dutt07, Balasubramanian08, Rittweger09, Maurer09, Gorshkov08}. The second level, termed a \emph{super-plaquette} ($\sim 10 \mu$m $\times 10 \mu$m), consists of a lattice of plaquettes whose computational registers are coupled through DSCBs.  At the highest level of the hierarchy, we consider an array of super-plaquettes, where individual super-plaquettes are controlled by confined microwave fields~\cite{Lee07}. In particular, micro-solenoids can confine fields to within super-plaquettes, allowing for parallel operations at the super-plaquette level. For example, as depicted in Fig.~2, independent microwave pulses can allow for simultaneous operations on the electronic spins of all computational NV registers within all super-plaquettes. In order to control registers at the super-plaquette boundaries, we define a dual super-plaquette lattice (Fig.~2a). Localized microwave fields within such a dual lattice can provide a smooth transition between the boundaries of neighboring super-plaquettes. 

\begin{figure}
\centering
\includegraphics[width=3.4in]{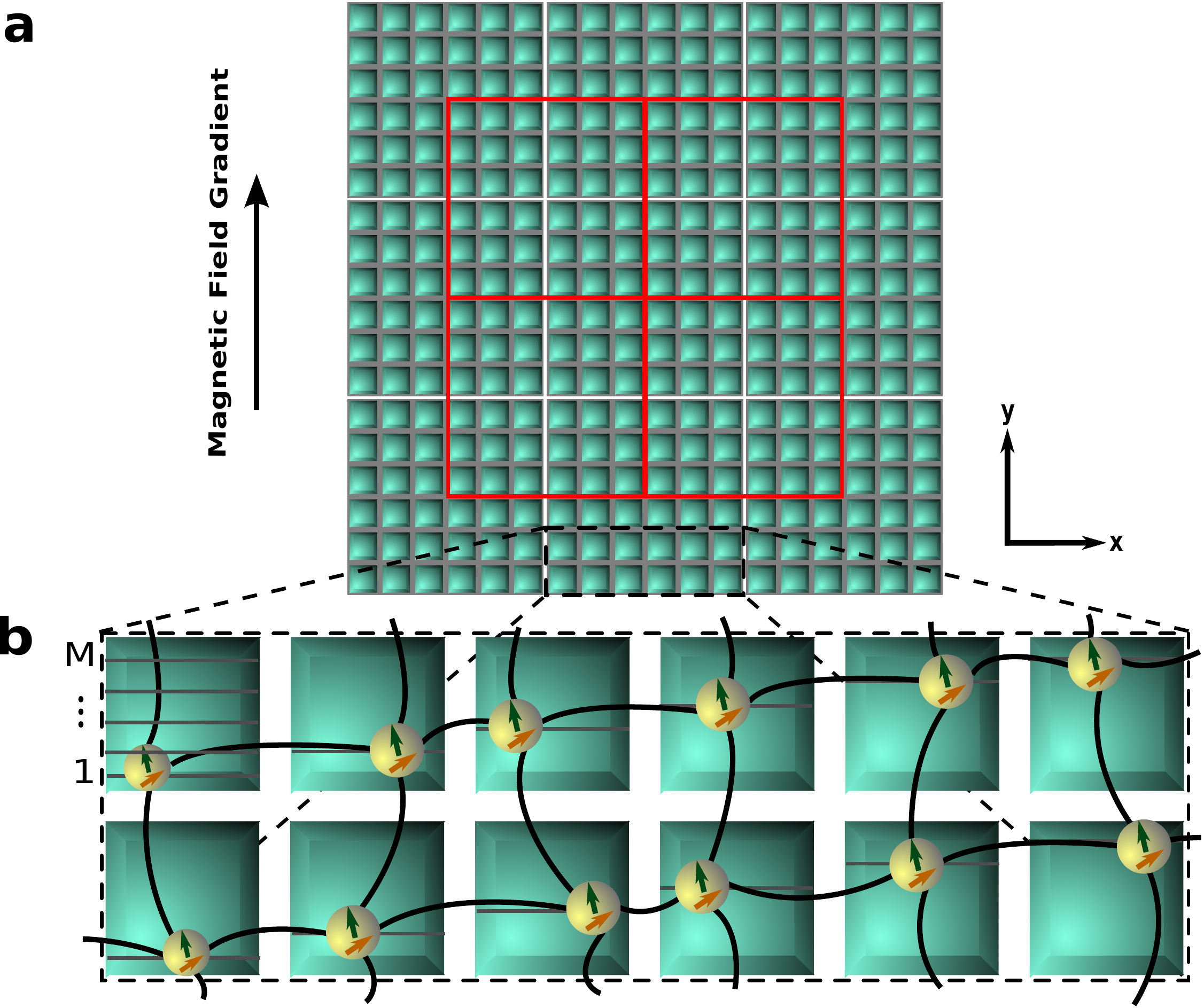}
\caption{The architecture for a room-temperature solid-state quantum computer.~(a) A two-dimensional hierarchical lattice allowing for length-scale based control, which enables fully parallel operations. At the lowest level, individual plaquettes are outlined in grey and each contains a single computational NV register. At the second level of hierarchy, a super-plaquette, outlined in white, encompasses a lattice of plaquettes; each super-plaquette is separately manipulated by micro-solenoid confined microwave fields. In order to allow for quantum information transfer across boundaries of super-plaquettes, there exists a dual super-plaquette lattice outlined in red.~(b) The schematic NV register implantation within a super-plaquette. Two rows of individual plaquettes within a super-plaquette are shown. NV registers, consisting of an electronic (green) and nuclear (yellow) spin are depicted within a staggered up-sloping array which is row-repetitive.  Individual rows within a \emph{single} plaquette are specified by an integer $n$ with $n=1$ being the bottom row and $n = M$ being the top row. To achieve a staggered structure, we specify a unique implantation row within each plaquette wherein single impurities are implanted and subsequently annealed. For a given row of plaquettes, the implantation row corresponding to the left-most plaquette is n = 1, while the plaquette to the immediate right has implantation row n = 2; this pattern continues until the final plaquette in a given row, which by construction, has the highest implantation row number. The implantation process is repeated for each row of plaquettes within the super-plaquette and creates an array of NV registers, which each occupy a unique row in the super-plaquette. Since each NV register occupies a unique row within the super-plaquette, the magnetic field gradient in the $\hat{y}$ direction allows for individual spectroscopic addressing of single registers. Coherent coupling of spatially separated NV registers in adjacent plaquettes is mediated by a dark spin chain data bus (DSCB) and is schematically represented by the curved line connecting individual registers (SI Text). The second implantation step corresponds to the creation of these horizontal and vertical dark spin chains.} 
 \end{figure}

Taking advantage of the separation of length scales inherent to optical control and microwave confinement provides a mechanism to achieve parallelism; indeed, the hierarchical control of plaquettes, super-plaquettes, and super-plaquette arrays allows for simultaneous single- and two-qubit gate operations, which are fundamental to fault-tolerant computation. One of the key difference in the currently proposed architecture as compared to previous proposals~\cite{Benjamin06, JTSL07b} is that the design here does not rely on optically resolved transitions, which are only accessible at cryogenic temperatures. 

The required 2D array of NV centers can be created via a two-step implantation process and the selective manipulation of individual registers within such an array is enabled by the application of a spatially dependent external magnetic field $B_{z}\left(y\right)  = \frac{dB_{z}}{dy}y+B_{z,0}$. The 1D magnetic field gradient is sufficiently strong to allow for spectroscopic microwave addressing of individual NV registers, each of which occupies a unique row in the super-plaquette, as shown in Fig.~2b. 

\section*{Results and Discussion}

\subsubsection*{Dark Spin Chain Data Bus}
To coherently couple two spatially separated NV centers, we consider two distinct approaches. First, we consider an approach, which is appropriate for spin-state transfer along the direction of the magnetic field gradient, in which individual addressing of spins is possible. This allows for an adiabatic sequential SWAP between neighboring qubits and, consequently, between the ends of the chain.  Alternatively, in the situation where individual addressing of spins is not possible (i.e.~direction transverse to the field gradient), we show that global control pulses achieve effective Hamiltonian evolution, which enables quantum state transfer through the spin chain. In both cases, we show that perfect state transfer and remote coupling gates are possible even when the intermediate spin chain is completely unpolarized (infinite spin temperature).

We begin by analyzing the adiabatic sequential SWAP in a spin-$1/2$ chain. This approach is suitable to couple registers in plaquettes that are vertically adjacent, relying upon the individual addressability of qubits and utilizing the magnetic dipole coupling between spin-chain elements. As shown in the SI Text, under the secular approximation, the magnetic dipole coupling between a pair of neighboring spins can be reduced to Ising form
\begin{equation}
H_{int} = 4\kappa S_{z}^{1}S_{z}^{2} + \sum_{i=1,2}(\omega_{0} + \delta_{i})S_{z}^{i},
\end{equation}
\noindent where $\kappa$ is the relevant component of the dipole tensor, $\omega_{0}$ captures the electronic Zeeman energy, and $\delta_{i}$ characterizes both the hyperfine term (nuclear spin dependent) and the magnetic field gradient (SI Text).  From the Ising Hamiltonian, an XX interaction between qubits can be distilled by driving with $H_{drive}=\sum_{i=1,2} 2\Omega_{i} S_{x}^{i}\cos[ (\omega_{0}+\delta_{i})t ]$, leading to (under the rotating wave approximation, in the rotating frame, and in a rotated basis with $(x,y,z) \rightarrow (z,-y,x)$)
\begin{equation}
H_{int} = \kappa (S_{1}^{+}S_{2}^{-} + S_{1}^{-}S_{2}^{+} ) + \Omega_{1} S_{z}^{1} + \Omega_{2} S_{z}^{2}.
\end{equation}

\noindent The spin-flip process in $H_{int}$ is highly suppressed in the limit of $|\Omega_{1} - \Omega_{2}| \gg \kappa$, while the same process is dominant in the case of $|\Omega_{1} - \Omega_{2}| \ll \kappa$. Hence, by slowly ramping the Rabi frequencies $\Omega_{1}$ and $\Omega_{2}$ through one another, adiabatic SWAP of the quantum states of the two impurities can be achieved through rapid adiabatic passage, as shown in Fig.~3a. Generalizing to arbitrary length spin chains yields $H_{int} = \sum_{i} \kappa (S_{i}^{+}S_{i+1}^{-} + S_{i}^{-}S_{i+1}^{+}) + \sum_{i} \Omega_{i} S_{z}^{i}$, whereby the sequential adiabatic SWAP of quantum states along the spin chain can be achieved by successively tuning individual Rabi frequencies across one another. During the adiabatic SWAP of a single pair of spins, higher order interactions, such as those resulting from next-to-nearest neighbors, will be suppressed due to the differences in Rabi frequencies. By including the magnetic dipole coupling between the electronic spin of the NV register and the spin chain quantum channel, we arrive at an effective mixed spin chain with the DSCB connecting the two electronic spins of the vertically separated NV registers. The specific procedure resulting in adiabatic sequential SWAP mediated coupling between NV registers is depicted in Fig.~3c.

\begin{figure}
\centering
\includegraphics[width=3.4in]{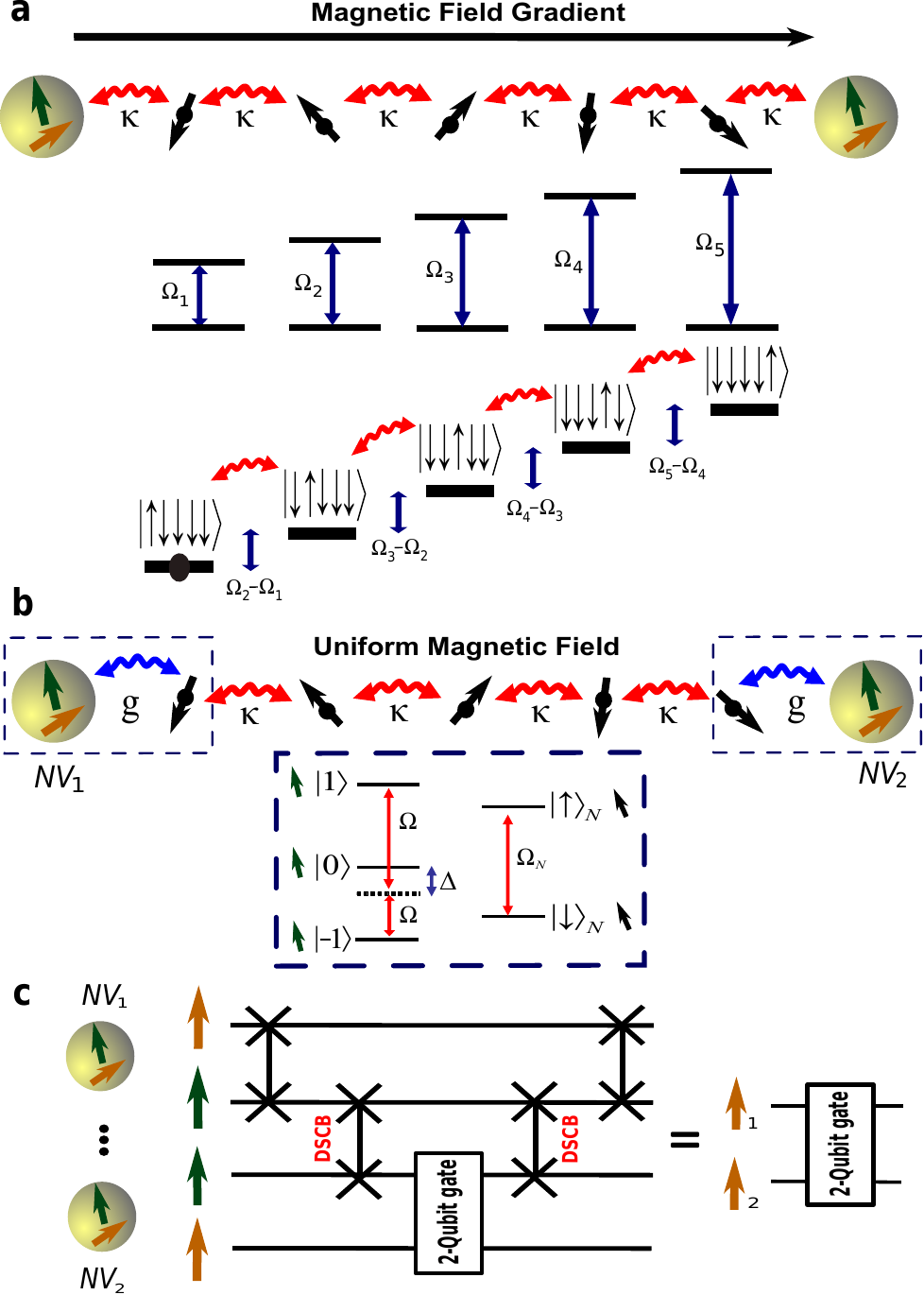}
\caption{Dark spin chain data bus (DSCB) mediated coherent coupling of spatially separated NV registers, which does not require spin chain initialization.~(a) Adiabatic sequential SWAP along the vertical direction, parallel to the magnetic field gradient. Individual addressing of impurities, enabled by the field gradient, allows for a slow ramping of the Rabi frequencies $\Omega_{i}$ and $\Omega_{j}$ through one another; this achieves adiabatic SWAP of the quantum states of the two impurities through rapid adiabatic passage. Thus, sequential adiabatic SWAP of quantum states along the spin chain can be achieved by successively tuning individual Rabi frequencies across one another.~(b) Free fermion state transfer in the horizontal direction, transverse to the magnetic field gradient. The coupling strength between the end qubits and the spin chain is $g$, while the interchain coupling strength is $\kappa$. Schematic representation of the level structure of the NV electronic spin and a dark impurity spin. Controlling the NV-impurity coupling $g$ is an essential component of FFST and occurs by driving the NV in two-photon resonance, with Rabi frequency $\Omega$ and detuning $\Delta$.~(c) Schematic circuit diagram outlining the protocol to achieve coherent coupling between the nuclear memory qubits of spatially separated NV registers. First, the nuclear and electronic qubits of a single register are swapped. Next, the electronic qubits of the two NV centers to be coupled are swapped via the DSCB. Finally, a two-qubit gate between the electronic and nuclear spin of the second register is performed before the memory qubit is returned to the nuclear spin of the original NV center. } 
\end{figure}

Crucially, such an adiabatic sequential SWAP is robust against variations in the coupling strength $\kappa$, which can be induced by the imprecise implantation of impurities that form the spin-$1/2$ chain; in particular, even for the case of varying $\kappa_{i,i+1}$, perfect adiabatic SWAP occurs so long as the rate at which $\Omega_{i}$ and $\Omega_{i+1}$ are ramped through one another is sufficiently small. Within the proposed architecture, the impurities forming the horizontal spin chain will not induce operational errors during the vertical adiabatic sequential SWAP since the design principle allows for selective spin echoing (SI Text).   


Next, we consider a second method, termed free fermion state transfer (FFST) developed in~\cite{YJG10}, to coherently couple NV registers. In contrast to the adiabatic sequential SWAP, the method utilizes only global control over impurities and effective Hamiltonian evolution. The relaxation of the requirement of individual control over elements of the dark spin chain renders this second method applicable for coherent coupling between NV registers in horizontally adjacent plaquettes, transverse to the direction of the field gradient. In particular, the protocol achieves coherent coupling through an unpolarized, infinite temperature spin chain, employing purely Hamiltonian evolution under
\begin{equation}
\begin{aligned}
H_{FFST} = &\hspace{1mm} g(S_{NV_{1}}^{+} S_{1}^{-} + S_{NV_{2}}^{+} S_{N}^{-} + \mbox{ h.c.}) \\
&+ \sum_{i=1}^{N-1} \kappa (S_{i}^{+} S_{i+1}^{-} + S_{i}^{-} S_{i+1}^{+})
\end{aligned}
\end{equation}
\noindent as shown in Fig.~3b. This Hamiltonian, obtained in analogy to Eq.~{\bf{3}}, results in coherent interactions between NV centers, which is best understood via an analogy to eigenmode tunneling in a many-body system. Specifically, the spin chain described by $H_{FFST}$ can viewed as a system of non-interacting fermions. As described in~\cite{YJG10}, by tuning the NV centers into resonance with a single fermionic eigenmode, an effective three-state system can be realized. Mediated by this fermionic eigenmode, the electronic states of two remote NV centers can be coherently swapped, leading to an analogous protocol for remote register coupling as shown in Fig.~3c. Crucially, such a SWAP gate is insensitive to the polarization of the intermediate dark spins and high-fidelity quantum state transfer can be achieved, provided that the fermionic mode is delocalized and that the coupling, $g$, of the NV qubit to the spin chain is controllable. As detailed in the Materials and Methods, by utilizing the three-level NV ground-state structure (Fig.~3b), it is possible to fully control the NV-chain coupling. This tunability also ensures that FFST is fundamentally robust to experimentally relevant coupling-strength disorder, which could be induced by implantation imprecision. Indeed, by separately tuning the NV-chain coupling on either side of the DSCB, it is possible to compensate for both disorder-induced asymmetry in the fermionic eigenmode as well as altered statistics of the eigenenergies \cite{YJG10, Evers08, Balents97}. The particular implementation of FFST within the NV architecture is further described in the SI Text.

\subsubsection*{Implementation, Operational Errors and Gate Fidelities} 

The specific implementation of the DSCB can be achieved with implanted Nitrogen impurity ions. Dipole coupling between neighboring Nitrogen electronic spins forms the DSCB, while dipole coupling between the NV and Nitrogen electronic spins forms the qubit-DSCB interaction; non-secular terms of this magnetic dipole coupling are highly suppressed due to the spatially dependent external magnetic field $B_{z}(y)$, resulting in the effective interaction found in Eq.~{\bf{2}}. In addition, the Nitrogen impurities possess a strong hyperfine coupling, the principal axis of which can take on four possible orientations due to tetrahedral symmetry~\cite{Takahashi08, Loubser67, Kedkaew08}. Dynamic Jahn-Teller (JT) reorientation of the Nitrogen impurity's hyperfine principal axis results in two particular considerations: 1) the addressing of additional JT frequencies yielding a denser super-plaquette frequency spectrum and 2) the JT-governed spin-lattice relaxation (SLR) time $T_{1}^{N}$ (SI Text). Since $T_{1}^{N}$ is characterized by an Arrhenius rate equation~\cite{Loubser67} at ambient temperatures, a combination of a static electric field and slight cooling by $\approx 50$K allows for a substantial extension of the relaxation time to $\sim 1$s; hence, in the following consideration of operational errors, we will assume that we are limited by $T_{1}^{NV}$, the spin-lattice relaxation time of the NV center.


\begin{figure}
\centering
\includegraphics[width=3.4in]{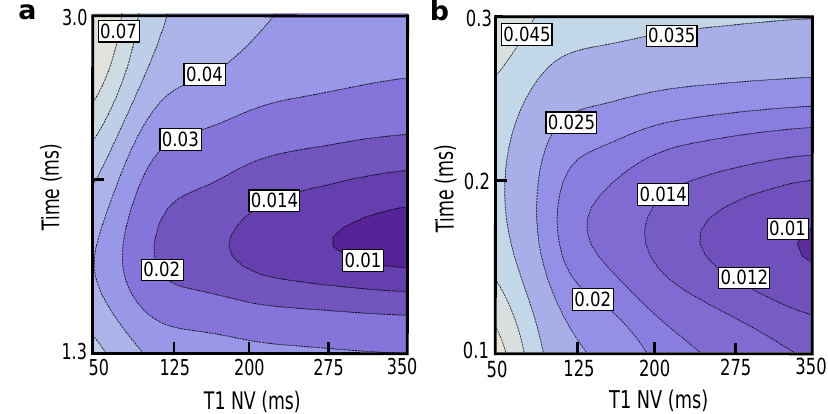}
\caption{Numerical simulation of the DSCB fidelity.~(a) The operational infidelity associated with the adiabatic sequential SWAP for $N=18$. The simulations account for the Jahn-Teller orientation of Nitrogen impurities and utilize the optimized adiabatic ramp profile~\cite{Roland02}. Simulations utilize an optimized coupling strength of $8.71$kHz ($18.1$nm spacing). Full numerical integration of the time dependent Schr\"odinger equation produces infidelity contour plots as a function of total SWAP time and $T_{1}^{NV}$.~(b) Numerical simulations of the operational infidelity associated with FFST for $N=7$. Non-nearest neighbor interactions are assumed to be refocused through dynamic decoupling as described in the SI Text. Simulations, which utilize an optimized coupling strength of $12.6$kHz ($16$nm spacing), are based upon a full diagonalization and also account for the Jahn-Teller orientation of Nitrogen impurities. Infidelity contour plots are again shown as a function of total SWAP time and $T_{1}^{NV}$.} 
\end{figure}

We now consider various imperfections, which may introduce operational errors. In particular, we consider the errors associated with: 1) sequential SWAP mediated coupling between vertically adjacent registers and 2) FFST between horizontally adjacent registers. We begin by discussing the analytic error estimate associated with each method, after which, we summarize the results of full numerical simulations. 

First, we consider the accumulated infidelity associated with the adiabatic sequential SWAP (SI Text),
\begin{equation}
p_{err}^{SS} \approx N (p_{off}^{SS} +p_{adia} +  p_{dip} + p_{T1}^{SS} + p_{T2}^{SS} ).
\end{equation}

\noindent The first term, $p_{off}^{SS} \sim  \left( \frac{\Omega_{i}}{\Delta_{g}}\right)^{2}$, represents off-resonant excitations induced by microwave manipulations with Rabi frequency $\Omega_{i}$. Here, $\Delta_{g}$ characterizes the gradient-induced splitting achieved within the super-plaquette frequency spectrum. The second term, $p_{adia}$, corresponds to the non-adiabatic correction resulting from an optimized adiabatic ramp profile~\cite{Roland02, Caneva09, Messiah62}. The third term, $p_{dip} \sim  \left( \frac{\kappa}{\Omega_{i}}\right)^{2}$, is directly obtained from Eq.~{\bf{3}} and corresponds to additional off-resonant errors. The fourth error term, $p_{T1}^{SS}$ corresponds to the depolarization error induced by the finite NV $T_{1}$ time, while the final error term, $p_{T2}^{SS}$ corresponds to the infidelity induced by dephasing. Since each error term is considered within the context of a single adiabatic SWAP, the total error contains an additional factor of $N$, representing the chain length, which is plaquette size dependent (e.g. $N \approx 5$ for $100$nm and $N\approx 20$ for $500$nm). 

We can similarly consider the accumulated infidelity associated with FFST (SI Text),
\begin{equation}
p_{err}^{FFST} \approx p_{off}^{FFST} + p_{fermi} + p_{g} + p_{T1}^{FFST}  + p_{T2}^{FFST}.
\end{equation}

\noindent In direct analogy to $p_{err}^{SS}$, the first term in $p_{err}^{FFST}$ corresponds to the excitation of an NV register by off-resonant microwave fields. The second term, $p_{fermi}$, corresponds to the undesired coupling with off-resonant fermionic modes. Since the coupling strength is characterized by $g/\sqrt{N}$~\cite{YJG10}, while the splitting of the eigenenergy spectrum $\sim \kappa/N$, such an off-resonant error induces an infidelity $\sim (\frac{g / \sqrt{N} }{\kappa / N})^{2}$. The third error term, $p_{g}$, results from the protocol designed to control, $g$, the NV-chain coupling (see Materials and Methods for details). Finally, directly analogous to $p_{err}^{SS}$, the fourth and fifth terms correspond to errors induced by the operational time, $t_{FFST}$, which causes both depolarization and dephasing.

Finally, we perform numerical simulations, taking into account the Nitrogen JT frequencies, to characterize the infidelity of both the adiabatic sequential SWAP and FFST within the NV architecture, as depicted in Fig.~4. The results of these calculations are in excellent agreement with the above theoretical predictions (SI Text). In particular, these simulations reveal that, for sufficiently long $T_{1}^{NV} \sim 100$ms, operational infidelities in both DSCB methods can be kept below $10^{-2}$. 

These simulations clearly show that the $T_{1}$ time of the NV electronic spin is of critical importance in obtaining high-fidelity quantum operations. While at room temperature $T_{1}$ appears to vary depending on the particular sample and on the specific properties of the local NV environment, such as strain, values on the order of $10$ms are generally obtained~\cite{Dutt07, Takahashi08}. However, the spin-lattice relaxation mechanism governing $T_{1}$ is most likely related to an Orbach process~\cite{Isoya10,Redman91}, which is strongly temperature dependent. In such a case, modest cooling of the sample by $\approx 50$K, is likely to extend T1 by more than an order of magnitude, thereby making high fidelity gates possible. 

Given that such numerical estimations suggest the possibility of achieving high fidelity two-qubit operations between remote NV registers, the proposed architecture seems well suited to the implementation of topological quantum error correction~\cite{Bravyi98, Raussendorf07, Fowler09, Wang10}. Recent progress in optimizing the 2D nearest-neighbor surface code has yielded an error threshold of $\epsilon \approx 1.4\%$~\cite{Wang10b}, which is above the estimated infidelity corresponding to both the adiabatic sequential SWAP and FFST; thus, in principle, implementation of the 2D surface code can allow for successful topological quantum error correction, and hence, fault tolerant quantum computation~\cite{Jones10}. 

\section*{Conclusions and Outlook}

The above considerations indicate the feasibility of experimentally realizing a solid-state quantum computer capable of operating under ambient conditions at or near room temperature. We emphasize that a majority of the elements required for the realization of individual qubits in our architecture have already been recently demonstrated. In our approach, these techniques are supplemented by both a new mechanism for remote register coupling between NV centers as well as a hierarchical design principle, which facilitates scalability. The remote coupling mechanisms discussed can naturally be implemented via Nitrogen ion implantation in ultra-pure diamond crystals and are robust to realistic imperfections and disorder~\cite{YJG10}. 

While the implementation and integration of the various proposed elements still require substantial advances in areas ranging from quantum control to materials science, a feasible approach to \emph{room temperature} quantum information processing can greatly alleviate the stringent requirements associated with cryogenic temperatures, thereby making the realization of a scalable quantum computer significantly more practical. 

The present work opens a number of new directions which can subsequently be explored. In particular, while we have considered the direct errors associated with DSCB mediated coupling, it is instructive to note that the fidelity of such quantum gates can often be significantly improved using techniques from optimal control theory~\cite{Khaneja01, Khaneja03, Yuan08}. For example, such methods of optimal control, while negating the detrimental effects of decoherence, can also simultaneously allow for the implementation of high-fidelity gates despite both frequency and coupling disorder as induced by ion implantation errors. Indeed, the ability to precisely guide the quantum evolution via optimal control, even when the system complexity is exacerbated by environmental coupling, provides an alternative solution to improve single and two-qubit gate fidelities~\cite{Grace07}. In addition, it is well known that the local strain field surrounding each NV center can significantly alter the register's properties; hence, through a detailed understanding of electric field induced strain, it may be possible to improve the coherence properties of the qubit. Beyond these specific applications, a number of scientific avenues can be explored, including for example, understanding and controlling the non-equilibrium dynamics of disordered spin systems.

\section*{Materials and Methods}
{\bf{Controlling Qubit-Chain Coupling in the NV Architecture}}--- To achieve an effective Hamiltonian of the form given by Eq. {\bf{4}}, it is essential to control the coupling strength between the NV register and the neighboring impurity. Here, we utilize the three levels of the NV electronic spin~\cite{Rabl09} to effectively control $g$, as shown in Fig.~3b, whereby the Hamiltonian (under microwave driving) can be written as
\begin{equation}
\begin{aligned}
H = & \hspace{1mm}-\Delta ( |1 \rangle \langle 1| + |-1 \rangle \langle -1|)\\
& - \Omega (|0 \rangle \langle 1| + |0 \rangle \langle -1| +\text{h.c.}) - \Omega_{N} S_{x}^{N} + 4\kappa S_{z}^{NV} S_{z}^{N},
\end{aligned}
\end{equation}
\noindent where $\Omega$ represents the Rabi frequency on the NV register, $\Delta$ represents the associated detuning, and $\Omega_{N}$ represents the Rabi frequency on the Nitrogen impurity. In this case, since the NV two-photon detuning is zero, it is convenient to define bright and dark states, $|B \rangle = \frac{| 1 \rangle + | -1 \rangle}{\sqrt{2}}$ and $|D \rangle = \frac{| 1 \rangle - | -1 \rangle}{\sqrt{2}}$; further, in the resulting two-level picture, the associated dressed states are $| + \rangle \approx | B \rangle + \frac{\sqrt{2} \Omega}{\Delta} | 0 \rangle$ and $| - \rangle \approx | 0 \rangle - \frac{\sqrt{2} \Omega}{\Delta} | B \rangle$, in the limit $\Omega \ll \Delta$. Hence, rewriting the Hamiltonian in this limit yields
\begin{equation}
\begin{aligned}
H = & \hspace{1mm} -\Delta |D \rangle \langle D| \hspace{1mm} - \hspace{1mm} (\Delta + \frac{2 \Omega^{2}}{\Delta}) |+ \rangle \langle +| \hspace{1mm}\\
& + \hspace{1mm} \frac{2 \Omega^{2}}{\Delta} |- \rangle \langle - | \hspace{1mm} - \hspace{1mm} \frac{1}{2} \Omega_{N} (|+\rangle_{N} \langle + | \hspace{1mm} - \hspace{1mm} |-\rangle_{N} \langle -|) \hspace{1mm} \\
&  + \hspace{1mm}2\kappa (|B \rangle \langle D| + |D \rangle \langle B|) (|+ \rangle_{N} \langle -| \hspace{1mm} + \hspace{1mm} |- \rangle_{N} \langle +|),
\end{aligned}
\end{equation}
\noindent where $|\pm \rangle_{N} = \frac{| \uparrow \rangle_{N} \pm | \downarrow \rangle_{N}}{\sqrt{2}} $ correspond to the two $S_{x}^{N}$-eigenstates of the Nitrogen impurity. The coupling term can be further re-expressed as
\begin{equation}
\begin{aligned}
2 \kappa & \left \{ (  |+ \rangle \langle D| + |D \rangle \langle +|) \hspace{1mm} - \frac{\sqrt{2} \Omega}{\Delta} \hspace{1mm} (|- \rangle \langle D| + |D \rangle \langle -|) \right \} \\
& * (|+_{N} \rangle \langle -_{N}| \hspace{1mm} + \hspace{1mm} |-_{N} \rangle \langle +_{N}|).
\end{aligned}
\end{equation}
\noindent  Thus, by working within the NV subspace $\{ | D \rangle , | - \rangle \}$, it is possible to completely control the coupling between the NV register and Nitrogen impurity, $g \sim \kappa \frac{\Omega}{\Delta}$, by tuning the Rabi frequency and detuning.  It is possible to work in the required two-state subspace by ensuring that $\kappa \ll \Delta$ and hence, that the $| + \rangle$ state remains unpopulated, with corresponding off-resonant error $\kappa^{2} / \Delta^{2}$. 

Furthermore, we evince a possible scheme to coherently map the quantum information that is stored in the nuclear memory into the desired electronic subspace. For example, consider mapping $ | 0 \rangle \otimes ( \alpha | \uparrow \rangle + \beta | \downarrow \rangle)$ to $(\alpha |- \rangle + \beta | D \rangle) \otimes | \uparrow \rangle$, where the first (tensor) factor corresponds to the electronic state and the second corresponds to the nuclear state of a single NV. The proposed mapping can be achieved in a two-step process. First, by simultaneously performing a $\pi-$pulse on the transitions $ | 0 \rangle \otimes| \downarrow \rangle \rightarrow  | -1 \rangle \otimes| \downarrow \rangle$  and $ | 0 \rangle \otimes| \downarrow \rangle \rightarrow  | 1 \rangle \otimes| \downarrow \rangle$ with oppositely signed Rabi frequencies, one can map $ | 0 \rangle \otimes| \downarrow \rangle$ to $ | D \rangle \otimes| \downarrow \rangle$. Next, one utilizes an RF pulse to flip the nuclear spin, which yields $ | D \rangle \otimes| \downarrow \rangle \rightarrow | D \rangle \otimes| \uparrow \rangle$. Finally, turning $\Omega$ on in an adiabatic fashion ensures that the state preparation populates only $|D \rangle$ and $| - \rangle$, thereby mapping the quantum information into the desired electronic subspace.

\section*{Acknowledgments}

We gratefully acknowledge conversations with G. Goldstein, J. Maze, E. Togan, Y. Chu, J. Otterbach, Z.-X. Gong, L.-M. Duan, C. Laumann, C. Mathy, A. Zhai, J. Preskill, N. Schuch, and Y. T. Siu.  This work was supported by the NSF, DOE (FG02-97ER25308), CUA, DARPA, AFOSR MURI, NIST, the DFG within SFB631 and the Nano Initiative Munich (NIM), the Lee A. DuBridge Fellowship and the Sherman Fairchild Foundation.

\clearpage

\begin{center}
\section*{{\large \bf{Supporting Information (SI Text)}}}
\end{center}
\subsection*{Realization of Dark Spin Chain Data Bus with Nitrogen Impurities}
\subsubsection*{Nitrogen Impurity Hamiltonian}

\noindent We consider a dipole-dipole coupled chain of nitrogen impurities which forms the basis for the DSCB that couples remote NV registers. The Hamiltonian for a single Nitrogen impurity is given by
\begin{equation*}
\begin{aligned}
H_{N} = & \hspace{1mm} g_{e} \mu_{B} \vec{B} \cdot \vec{S} - g_{N} \mu_{N} \vec{B} \cdot \vec{I}  \\
& + A_{\parallel} S_{z'}I_{z'} + A_{\perp} (S_{x'}I_{x'}+S_{y'}I_{y'}), 
\end{aligned}
\tag{S1}
\end{equation*}
\noindent where $A_{\parallel} = -159.7$MHz and $A_{\perp} = -113.8$MHz are the hyperfine constants corresponding to the primed axes, which is chosen with the Jahn-Teller (JT) axis as $z'$~\cite{Kedkaew08X}. As in recent reports~\cite{Maurer09X}, we consider the application of a magnetic field $\vec{B}$ of strength $B_{z}$ ($\sim 1$T), as defined in the main text, throughout the 2D array along the NV axis (consider e.g. $\hat{NV} \parallel (1,1,1)$), which defines the quantization axis $\hat{z}$. To model the additional Jahn-Teller frequencies applied in the super-plaquettes, we calculate the impurity chain Hamiltonian between two nitrogens with differing Jahn-Teller axes (assuming a $(1,1,1)$ cut diamond, as shown in Fig.~S1), and with the JT axis of $N_{1}$ parallel to the NV axis (note that the non-parallel JT axes are equivalent to one another relative to the NV axis). We explicitly represent the hyperfine interaction within a standard Cartesian basis by choosing the Jahn-Teller axis of the first nitrogen $N_{1}$ as $e_{z'} \parallel (1,1,1)$ and subsequently $e_{x'} \parallel (2,-1,-1)$ and $e_{y'} \parallel (0,1,-1)$, and of the second nitrogen $N_{2}$ as $e_{z'} \parallel (1,1,-1)$ and subsequently $e_{x'} \parallel (-2,1,-1)$ and $e_{y'} \parallel (0,1,1)$ (the generalization to other pairs of JT axes follows after). Here, the hyperfine interaction of $N_{1}$ then takes the form $H_{HF} = A_{\parallel} S_{z}I_{z} + A_{\perp} (S_{x}I_{x}+S_{y}I_{y})$, while the hyperfine term of the $N_{2}$ takes the form 
$\frac{1}{3} A_{\parallel} (\frac{S_{z}}{\sqrt{3}} + \frac{2S_{x}}{\sqrt{6}} + \frac{2S_{y}}{\sqrt{2}})(\frac{I_{z}}{\sqrt{3}} + \frac{2I_{x}}{\sqrt{6}} + \frac{2I_{y}}{\sqrt{2}}) + A_{\perp} [ \frac{1}{6} (\frac{-2S_{z}}{\sqrt{3}} + \frac{2S_{y}}{\sqrt{2}} -\frac{4S_{x}}{\sqrt{6}})(\frac{-2I_{z}}{\sqrt{3}} + \frac{2I_{y}}{\sqrt{2}} -\frac{4I_{x}}{\sqrt{6}}) 
+ \frac{1}{2} ( \frac{2S_{z}}{\sqrt{3}} -\frac{2S_{x}}{\sqrt{6}})(\frac{2I_{z}}{\sqrt{3}} -\frac{2I_{x}}{\sqrt{6}})] $. It is important to note that the electronic Zeeman term in the impurity Hamiltonian provides a shift $\sim 10$GHz and that the nuclear Zeeman term provides a shift $\sim 1$MHz. Thus, to good approximation, we can disregard electronic spin flip terms proportional to $S_{x}$ and $S_{y}$ since the hyperfine constants are two orders of magnitude smaller than the electronic Zeeman energy, yielding $H_{HF} \approx A_{\parallel} S_{z}I_{z}$ for $N_{1}$ and 
\begin{equation*}
\begin{aligned}
H_{HF} \approx & \hspace{1mm}S_{z}I_{z} (\frac{A_{\parallel}}{9} + \frac{8  A_{\perp}}{9}) + S_{z} (\frac{A_{\parallel}}{3\sqrt{3}} - \frac{  A_{\perp}}{3\sqrt{3}})(\frac{2I_{x}}{\sqrt{6}}+\frac{2I_{y}}{\sqrt{2}}) \\
&  \equiv S_{z} (\gamma I_{z} + \alpha I_{x} + \beta I_{y}) 
\end{aligned}
\tag{S2}
\end{equation*}
\noindent for $N_{2}$, where $(\alpha,\beta,\gamma) = (-7.2,-12.5,-118.9)$MHz. For $N_{2}$, the hyperfine subspace is separated into two manifolds of nuclear spin $1/2$ and $-1/2$; to see that the nuclear spin-flip terms are highly off-resonant, it is easiest to work in a rotated basis with $\hat{\tilde{z}} = \frac{1}{\sqrt{\alpha^{2}+\beta^{2}+\gamma^{2}}}( \alpha \hat{x} +\beta \hat{y} + \gamma \hat{z}) $, $\hat{\tilde{x}} =   \frac{1}{\sqrt{\alpha^{2}+\gamma^{2}}}( -\gamma \hat{x}  + \alpha \hat{z})$, and $\hat{\tilde{y}} =   \frac{\beta \gamma}{\sqrt{(\alpha^{2}+\gamma^{2})(\alpha^{2}+\beta^{2}+\gamma^{2})}}( \frac{\alpha}{\gamma} \hat{x}  -\frac{1}{\beta}(\gamma+\frac{\alpha^{2}}{\gamma}) \hat{y} + \hat{z})$, wherein $H_{HF} \approx \sqrt{\alpha^{2}+\beta^{2}+\gamma^{2}} S_{z}I_{\tilde{z}}$. In the rotated basis, nuclear spin flips are captured by the Zeeman term $- g_{N} \mu_{N} B_{0}I_{z} = - g_{N} \mu_{N} B_{0} ( \frac{\gamma}{ \sqrt{\alpha^{2}+\beta^{2}+\gamma^{2}}}I_{\tilde{z}}+ \frac{\alpha}{ \sqrt{\alpha^{2}+\gamma^{2}}}I_{\tilde{x}}+\frac{\beta \gamma}{\sqrt{(\alpha^{2}+\gamma^{2})(\alpha^{2}+\beta^{2}+\gamma^{2})}}I_{\tilde{y}}  )$. Thus, for a given electronic spin, the nuclear spin subspaces are separated by $ \sqrt{\alpha^{2}+\beta^{2}+\gamma^{2}} \sim 100$MHz, while the strength of the nuclear spin flip terms proportional to $I_{\tilde{x}}$ and $I_{\tilde{y}}$ $\sim 100$kHz and can hence be neglected. These secular approximations yield a nearest neighbor Hamiltonian given by (returning to the original basis for simplicity of notation)
\begin{equation}
\begin{aligned}
H_{nn} = &\hspace{1mm} g_{e} \mu_{B} B_{0}(S_{z}^{1}+S_{z}^{2}) - g_{N} \mu_{N} B_{0} (I_{z}^{1} +I_{z}^{2})\\
&  + A_{\parallel} S_{z}^{1}I_{z}^{1} + S_{z}^{2}I_{z}^{2} (\frac{A_{\parallel}}{9} + \frac{8  A_{\perp}}{9}) \\
& + \frac{ \mu_{0}}{4 \pi r^{3}} g_{e}^{2} \mu_{B}^{2} (S_{z}^{1}S_{z}^{2}) 
\end{aligned}
\tag{S3}
\end{equation}
\noindent where we have additionally neglected electronic spin flip terms originating from dipole-dipole interactions between the impurities since their hyperfine terms vary by $\sim 10$MHz while the dipole coupling strength is only $\sim 10$kHz for $20$nm separation (neglecting the field gradient). Thus, in the case where neighboring impurities have differing Jahn-Teller axes, the electronic spin flip terms can always be neglected to give an Ising interaction. In the case of parallel Jahn-Teller axes, when the neighboring nitrogen nuclear spins are different, the combined hyperfine term takes the form $\pm(S_{z}^{1}-S_{z}^{2}) (\frac{A_{\parallel}}{6} + \frac{A_{\perp}}{3})$. The dipole-dipole terms which correspond to electronic spin flipping will attempt to couple $ | \uparrow_{1}, \downarrow_{2} \rangle$ and $ | \downarrow_{1}, \uparrow_{2} \rangle$; however, this coupling is again highly suppressed since these states are separated by $(\frac{A_{\parallel}}{3} + \frac{2A_{\perp}}{3})\sim 100$MHz, while the dipole coupling strength is $\sim 10$kHz. Furthermore, the states $ | \uparrow_{1}, \uparrow_{2} \rangle$ and $ | \downarrow_{1}, \downarrow_{2} \rangle$ are separated from the others by the electronic Zeeman energy, ensuring that dipole-dipole induced spin flips will again be highly off-resonant. Finally, when the Jahn-Teller axes are parallel and the nuclear spins are also identical, these dipolar spin-flip terms are suppressed by the external magnetic field gradient.  It is important to note that for each Nitrogen spin, since the hyperfine term can take on four possible values, all four frequencies, corresponding to $\omega_{0} \pm \frac{1}{2}  A_{\parallel}$ and $\omega_{0}\pm \frac{1}{2} (\frac{A_{\parallel}}{9} + \frac{8  A_{\perp}}{9}) $ are applied in order to address the impurity ($\omega_{0}$ as defined in Eq. {\bf{2}} of the main text). 

\begin{figure}[t]
  \begin{center}
   \includegraphics[scale=1.2]{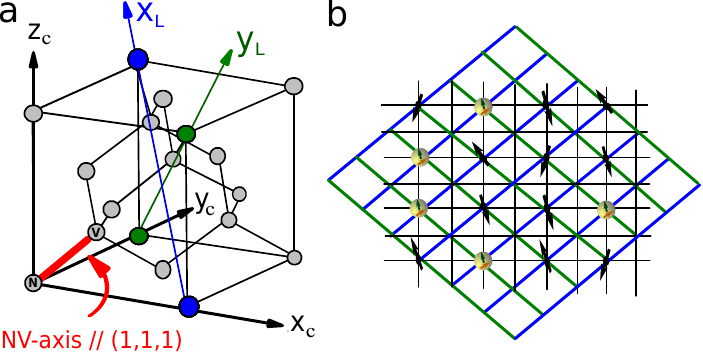}
  \end{center}
    \begin{flushleft}
\noindent \textbf{FIG.~S1:} NV Schematic.~(a) Depicts the diamond unit cell with corresponding NV axis $\parallel (1,1,1)$. The underlying lattice of the 2D array can be chosen with $x_{L}$ and $y_{L}$ defining the lattice grid (where $(x_{c},y_{c},z_{c})$ represent the Cartesian basis).  For simplicity, in the derivations, we assume a $(1,1,1)$-cut diamond, whereby the plane of the cut is parallel to the computational plane spanned by $x_{L}$ and $y_{L}$.~(b) In this particular case, all NV centers and Nitrogen impurities sit in the computational plane defined by $x_{L}$, $y_{L}$. We note that this choice of axes is in slight contrast to the main text, where for simplicity $\hat{x}$, $\hat{y}$ is chosen to represent the lattice in which all spins lie. Dark spins represent Nitrogen impurities while NV centers are represented as pairs of green (electronic) and yellow (nuclear) spins (in analogy to the main text). Here, in contrast to the sparsely occupied plaquettes in the main text, we have chosen to occupy all potential sites in the computational plane with spins for illustrative purposes.
\end{flushleft}
\end{figure}

\subsubsection*{Dynamic Jahn-Teller Re-orientation}

As mentioned in the main text, it is essential to consider the intrinsic properties of the dark Nitrogen impurity spin chain; in particular, the room temperature $T_{1}^{N} \approx 2$ms time of the Nitrogen impurity is limited by dynamic Jahn-Teller reorientation~\cite{Reynhardt98X, Zaritskii76X, Terblanche00X}. Thus, nominally, the total time required for both DSCB mediated coherent coupling protocols introduces a significant depolarization error if we consider limitation by the impurity $T_{1}^{N}$ time. However, at room temperature, the dynamic Jahn-Teller (JT) reorientation is governed by tunneling between the four tetrahedrally symmetric axes, and the associated rate is given by~\cite{Loubser67X} 
\begin{equation}
1/T_{1}^{N} = 4*10^{12} e^{-\epsilon /kT} \textrm{s}^{-1}, \tag{S4}\\
\end{equation}

\noindent where $\epsilon = 0.76$eV is the experimentally determined activation energy. This exponential dependence on both activation energy and temperature suggests that a combination of a static electric field ($\sim 10^6$V/cm)~\cite{Averkiev98X, Trew91X, Liu78X} and slight cooling ($T \approx 250$K) can extend $T_{1}^{N}$ to above $10$s. Hence, in the discussion of operational errors in the main text, we have assumed limitation by $T_{1}^{NV}$, the spin-lattice relaxation time of the NV center.

\subsubsection*{Interaction between NV Qubit and N Impurity}

\noindent In addition to the properties associated with single Nitrogen impurities, we now consider the NV-N interaction and show that it is possible to distill an Ising interaction. The NV Hamiltonian is given by
\begin{equation*}
\begin{aligned}
H_{NV} = & \hspace{1mm} g_{e} \mu_{B} B_{z} S_{z} - g_{N} \mu_{N} B_{z} I_{z}\\
&  + A_{NV} (S_{z}I_{z} +S_{x}I_{x}+S_{y}I_{y}) +\Delta_{0} S_{z}^{2}
\end{aligned}
\tag{S5}
\end{equation*}
\noindent where $A_{NV} = 3.1$MHz and $\Delta_{0} = 2.87$GHz is the zero-field splitting of the singlet~\cite{Childress06X}. For the purposes of this derivation, we will consider the $m_{s} =0,1$ states to be the NV electronic qubit and will disregard the $m_{s} = -1$ state which is far-detuned (Zeeman shifted away due to the $\sim 1$T magnetic field). The flip terms between the electronic and nuclear spin ($S^{+}I^{-} +S^{-}I^{+}$) of the NV are suppressed due to the zero-field splitting and the large magnetic field, since $\Delta_{0}, g_{e}\mu_{B}B_{z} \gg A_{NV}$. The dipole-dipole interactions between the NV electronic spin and an impurity electronic spin will be nearly identical to the impurity-impurity dipole coupling which has previously been considered. In particular, considering quantization along the NV axis, which for simplicity will also be the assumed Jahn-Teller axis (the result is analogous for other JT orientations), and noting that the unit vector pointing from the NV to the N-impurity lies in the computational plane defined by the normal vector $(1,1,1)$, we find that the Hamiltonian is (under the secular approximation in analogy to Sec. IA),
\begin{equation*}
\begin{aligned}
H = & \hspace{1mm} g_{e} \mu_{B} B_{z} (S_{z}^{NV} + S_{z}^{N}) - g_{N} \mu_{N} B_{z} (I_{z}^{NV} + I_{z}^{N})\\
&  + A_{NV} S_{z}^{NV}I_{z}^{NV}+\Delta_{0} (S_{z}^{NV})^{2}\\
& + A_{\parallel} S_{z}^{N}I_{z}^{N} + \frac{ \mu_{0}}{4 \pi r^{3}} g^{2} \mu_{B}^{2} S_{z}^{NV}S_{z}^{N} 
\end{aligned}
\tag{S6}
\end{equation*}
\noindent To keep only the Ising interaction term, we require $\omega_{NV}$, $\omega_{N}$, and $|\omega_{NV}-\omega_{N}|$ to all be much greater than the strength of the dipole coupling, where $\omega_{NV}$ is the coefficient of $S_{z}^{NV}$ and $\omega_{N}$ is the coefficient of $S_{z}^{N}$.

\subsubsection*{Simulations and Error Optimization} 

\noindent We consider the optimization of errors in the context of the DSCB mediated remote register coupling.  First, we consider the accumulated infidelity associated with the adiabatic SWAP sequence, represented by Eq. {\bf{5}} in the main text, $p_{err}^{SS} \approx N ( p_{adia} + p_{off}^{SS} + p_{dip} + p_{T1}^{SS} + p_{T2}^{SS} )$. Re-expressing $p_{err}^{SS}$ to capture the form of the associated infidelity yields
\begin{equation}
p_{err}^{SS} \approx N \left [ \frac{1}{(\kappa' t_{ss})^{2}} +  \frac{\Omega_{i}^{2} }{ \Delta_{g}^{2}}+ \left( \frac{\kappa'}{\Omega_{i}}\right)^{2} + \frac{t_{ss}}{T_{1}^{NV}}  + \left( \frac{t_{ss}}{T_{2}^{NV}}\right)^{3} \right ]~, 
\tag{S7}
\end{equation}

\noindent where $t_{ss}$ represents the time required to SWAP between a single pair of impurities, $\kappa' \approx  8.7$kHz, $\Omega_{i}$ represents the Rabi frequency on the impurities, $\Delta_{g} = 10$MHz represents the gradient-induced splitting for a magnetic field gradient $\sim 10^{5}$T$/$m, $T_{1}^{NV} = 250$ms, and $T_{2}^{NV}=10$ms (see Sec. IIC for details on $\Delta_{g}$ derivation). Here, we describe the error terms in more detail to evince the origin of the infidelity. The first term in $p_{err}^{SS}$ corresponds to the non-adiabatic correction resulting from an optimized adiabatic ramp profile~\cite{Roland02X, Caneva09X, Messiah62X}. The second term represents off-resonant excitations induced by $\Omega_{i}$. The third term corresponds to additional off-resonant errors induced by the finite initial splitting as well as by dipole-dipole coupling of next-to-nearest neighbor impurities; since the dipolar interaction strength $\sim \kappa'$ while the characteristic energy spacing between impurities $\sim \Omega_{i}$, the associated error is $ \left( \frac{\kappa'}{\Omega_{i}}\right)^{2}$. The fourth error term corresponds to the depolarization error induced by the finite NV $T_{1}$ time, $t_{ss}/T_{1}^{NV}$. The final error term corresponds to the infidelity induced by dephasing $1-e^{-\left(t_{ss}/T_{2}^{NV}\right)^{3}} \approx (\frac{t_{ss}}{T_{2}^{NV}})^{3}$~\cite{Childress06X, Dutt07X}. 

The combined total error can be numerically optimized. However, before discussing the results of this optimization, we note that the limiting time scale in the system corresponds to $T_{1}^{NV}$; in particular, error estimates and numerical optimization are nearly independent of $T_{2}^{NV}$, so long as  $T_{2}^{NV} \gtrsim10$ms, which is experimentally achievable in isotopically purified diamond samples~\cite{Balasubramanian09X} . Furthermore, we note that the current form of $p_{T2}$ assumes that only a single echo pulse is applied during an individual SWAP. This suggests that the  functional form of the infidelity can be significantly improved by increasing the number of echo pulses, which is in effect already achieved in our system due to the assumed strong driving~\cite{Lange10X, Taylor06X}. Hence, in the remaining error analysis, we assume that dephasing-induced errors can be neglected. Numerically optimizing the full infidelity $p_{err}^{SS}$ in parameters $t_{ss}$ and $\Omega_{i}$ (for $N=18$), yields $\Omega_{i} \approx 450$kHz and $Nt_{ss} \approx 3$ms with total error $\approx 2.6*10^{-2}$, in good agreement with the full numerical simulations presented in the main text. The numerical simulations account for the additional Jahn-Teller (MW) frequencies and are obtained through numerical integration of the Schr\"odinger equation with the optimized adiabatic SWAP profile. Non-unitary errors corresponding to depolarization are then added to the unitary errors and the total infidelity is subsequently optimized. 

Next, we consider the accumulated infidelity associated with FFST, represented by Eq.~{\bf{6}} in the main text $p_{err}^{FFST} \approx p_{off}^{FFST} + p_{fermi} + p_{g} + p_{T1}^{FFST} $ (neglecting $p_{T2}^{FFST}$ as discussed above). Re-expressing $p_{err}^{FFST}$ to capture the form of the associated infidelity yields
\begin{equation}
p_{err}^{FFST} \approx \frac{\Omega_{N}^{2} + \Omega^{2}}{\left(  \Delta_{g} \right)^{2}} +(\frac{g / \sqrt{N} }{\kappa / N})^{2} + (\frac{\kappa}{\Delta})^{2}+  N\frac{t_{FFST}}{ T_{1}^{NV}},  \tag{S8}\\
\end{equation}

\noindent where $\Omega_N$ corresponds to the Rabi frequency applied on the impurity, $\Omega$ corresponds to the Rabi frequency applied on the NV register, $\kappa \approx  12.6$kHz, $t_{FFST}$ represents the total time required for state transfer, and $\Delta$ corresponds to the NV detuning as defined in the Methods section in the main text. Here, we describe the error terms in more detail to evince the origin of the infidelity.  The first term in $p_{err}^{FFST}$ corresponds to the off-resonant excitation of an NV register.  Similarly, the second term also results from an off-resonant error and corresponds to undesired coupling with off-resonant fermionic modes~\cite{YJG10X}. Since the coupling strength is characterized by $g/\sqrt{N}$, while the characteristic eigenenergy splitting $\sim \kappa/N$, such an off-resonant error induces an infidelity $(\frac{g / \sqrt{N} }{\kappa / N})^{2}$. The third error term results from the protocol designed to achieve controlled coupling $g$, as elucidated in the Methods. Finally, directly analogous to $p_{err}^{SS}$, the final term corresponds to the error induced by the operational time, $t_{FFST} \sim \sqrt{N}/g$, which results in depolarization. However, it is essential to note that relative to the adiabatic sequential SWAP, there is an additional factor of $N$ in $p_{T1}^{FFST}$. 

This factor results from the generation of multi-partite entanglement via a set of controlled-phase gates during a single FFST step; despite such entanglement, remote register coupling is readily achieved because a second transfer step (corresponding to the return of the quantum information to the original NV as shown in Fig.~3C of the main text) disentangles the quantum information from the intermediate dark spin chain data bus~\cite{YJG10X}. Crucially, the local two-qubit gate which is performed in the middle of two FFST steps must commute with the controlled-phase multi-partite entanglement. In particular, the two-qubit operation depicted in Fig.~3C could be a local CP-gate, which would be preserved in the disentangling step. Combined with single qubit rotations, a controlled-phase gate allows for arbitrary two-qubit gates and hence universal computation.

The controlled coupling $g$ is achieved by utilizing the NV three-level structure, whereby $g \sim \kappa \frac{\Omega}{\Delta}$ (as discussed in Methods section of the main text), allowing for the re-expression of $g$ in $p_{err}^{FFST}$. Similarly, since FFST requires tuning to a particular fermionic eigenmode, it is also possible to re-express $\Delta \approx \Omega_{N}$ (see Sec. IIB for details on fermion tuning). These re-expressions allow for numerical optimization in the two parameters $\Omega_{N}$ and $\Omega$ yielding $\Omega_{N} \sim 285$kHz and $\Omega \sim 95$kHz, with total operational errors $\approx 2.4*10^{-2}$ and total transfer time $t_{FFST} = 0.21$ms (where we have utilized parameters: $T_{1}^{NV}=250$ms and $N=7$), in good agreement with the full numerical simulations presented in the main text. The numerical simulations of FFST also account for the additional Jahn-Teller frequencies and are obtained through full Hamiltonian diagonalization. Non-unitary errors corresponding to depolarization are then added and the total infidelity is subsequently optimized. 

\begin{figure}[!ht]
  \begin{center}
   \includegraphics[scale = 0.59]{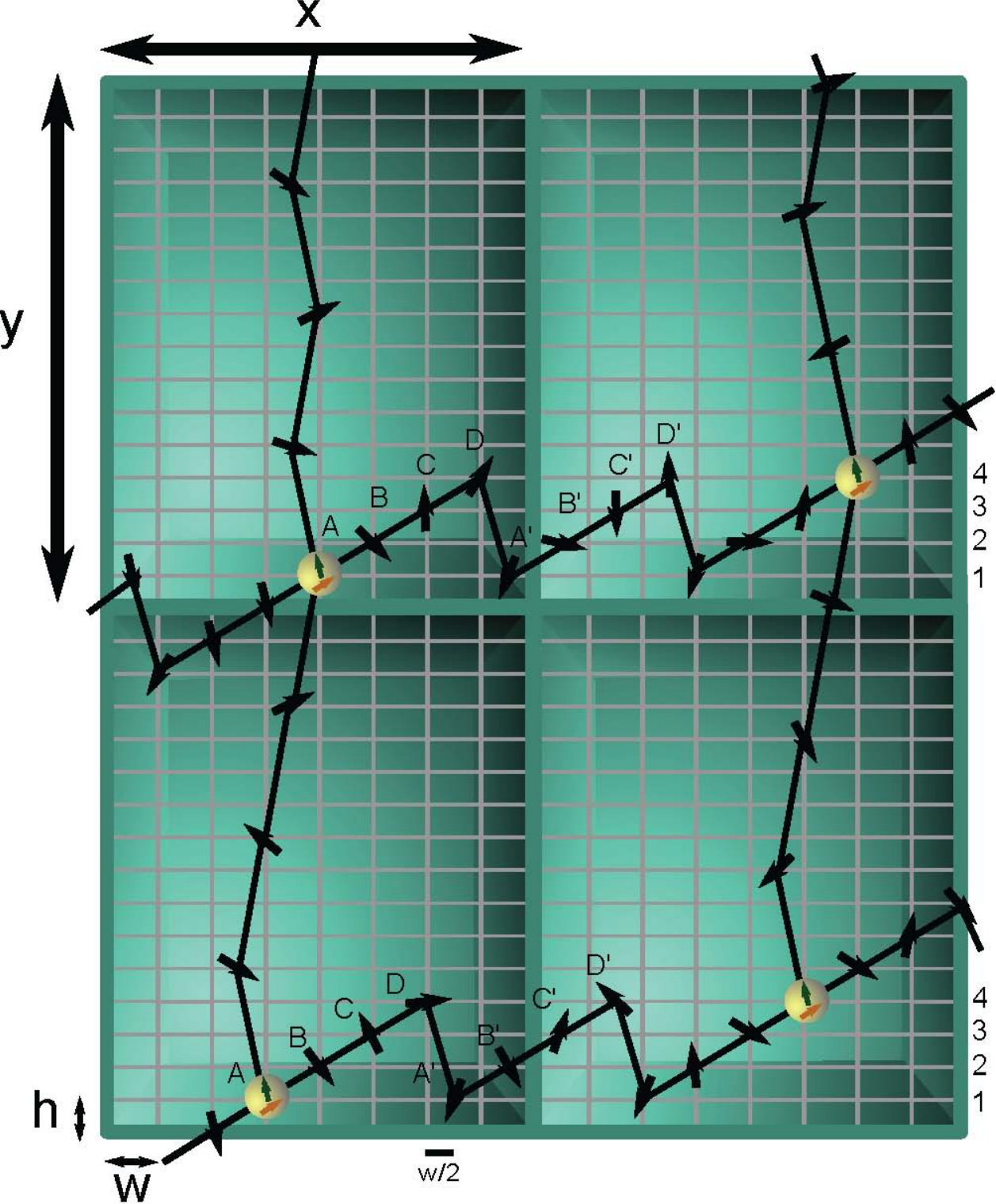}
  \end{center}
    \begin{flushleft}
\noindent \textbf{FIG.~S2:} \footnotesize Depicts the design principles governing the architecture. Specific impurity implantation within the architecture, which is designed to achieve parallel gate operations. Plaquettes, outlined in grey, are chosen with $y \approx 650$nm and $x \approx 525$nm; the individual plaquette-lattice spacing is chosen with $h \approx 6$nm and $w \approx 19$nm. The distance between all nearest neighbor pairs in the horizontal direction is $20$nm ($\sqrt{h^{2}+w^{2}} =$ $ \sqrt{(3h)^{2}+(w/2)^{2}}$) and the $\hat{x}$ spacing between spin $D$ and spin $A'$ is given by $w/2$. We note that the figure is not drawn to scale and that all impurity spins sit on the horizontal lattice, while some impurity spins sit at midpoints of the vertical lattice. The particular lattice spacing is chosen to ensure that the spatial separation between individual Nitrogen impurities forming the spin chain, as well as between NV registers and Nitrogens does not exceed $\approx 20$nm. The chosen plaquette-lattice spacing implies that neighboring plaquettes are coupled by $\approx 25$ Nitrogen impurities (black spins) in the $\hat{x}$ direction and by $\approx 25$ Nitrogen impurities in the $\hat{y}$ direction. The zig-zag pattern of impurity implantation in the $\hat{y}$ direction ($w/2$ horizontal distance) ensures that all spin chain links are of identical length, thereby allowing for parallel sequential SWAP operations along $\hat{y}$.  Additionally, such a pattern overcomes the limitation of imperfect NV conversion efficiency by allowing each vertical spin chain to span a sufficiently large horizontal spacing. Similarly, the saw-tooth pattern of impurity implantation in the $\hat{x}$ direction is designed to allow for echo pulse sequences which enable the cancelling of next-to-nearest neighbor interactions. In addition to refocusing next-to-nearest neighbor interactions, since the impurities forming the horizontal spin chain occupy unique rows relative to the vertical spin chain, during the adiabatic sequential SWAP, selective spin echoing of the horizontal chain impurities ensures that dipole coupling from such impurities will not induce operational errors during the vertical SWAP sequence. Similarly, selective spin echoing of vertical spin chain impurities during FFST can also suppress analogous operational errors. Finally, while the architecture is designed to showcase the ability to utilize the adiabatic sequential SWAP and FFST in differing directions, it is important to note that within the specific blueprint above, either implementation of the DSCB can be used in both directions. 
\end{flushleft}
\end{figure}

\subsection*{Specific Implementation of Architecture}
\noindent Here, we offer a specific implementation of the architectural design principle and discuss the various ingredients required to achieve DSCB-mediated coherent coupling between spatially separated NV registers. In particular, we consider the refocusing of non-nearest neighbor interactions along the horizontal spin chain, the control of directionality during FFST, and the full super-plaquette frequency spectrum.

\subsubsection*{Refocusing Next-Nearest-Neighbor Interactions}

\noindent The effective Hamiltonian evinced in Eq.~{\bf{4}} of the main text has nearest neighbor form; although next-nearest-neighbor interactions will represent a correction, we show that such interactions can be refocused within the current architecture design, and further, that in principle, interactions beyond next-nearest-neighbor can also be refocused. In particular, the horizontal spin chain ($N$ total spins) is arranged in a staggered saw-tooth fashion, as shown in Fig.~S2. Within such an architecture, nearest neighbor coupling terms correspond to all pairs of adjacent spins, each separated by $\approx 20$nm, with corresponding interaction Hamiltonian
\begin{equation}
H_{N}= \sum \kappa_{N} ( S_{A}^{+}S_{B}^{-} + S_{B}^{+}S_{C}^{-} +S_{C}^{+}S_{D}^{-} +S_{D}^{+}S_{A'}^{-} ) + \text{h.c.} \tag{S9}\\
\end{equation}

\noindent where the sum runs over all nearest neighbor pairs in a given dark spin chain. Thus, next-to-nearest neighbor terms for each spin correspond to the subsequent strongest interaction
\begin{equation*}
\begin{aligned}
H_{NN}= & \hspace{1mm} \sum \kappa_{NN} (S_{A}^{+}S_{C}^{-}+ S_{B}^{+}S_{D}^{-})\\
& + \kappa_{NN'} (S_{C}^{+}S_{A'}^{-}+ S_{D}^{+}S_{B'}^{-}) + \text{h.c.} 
\end{aligned}
\tag{S10}
\end{equation*}
\noindent where the prime denotes the next link in the saw-tooth chain as shown in Fig.~S2. In addition to the impurity spins, FFST incorporates the electronic spin of the NV register into a mixed spin chain. It is important to note that the spin-flip Hamiltonians $H_{N}$ and $H_{NN}$ are derived from the secular approximated Ising coupling by the application of driving fields as per Eq. {\bf{3}} in the main text. Since each row ($1,2,3,4$) is separately addressable by virtue of the magnetic field gradient (again, applying four frequencies for each row to ensure that all JT and nuclear spin states are addressed), it is possible to apply a spin-echo procedure to refocus the next-nearest-neighbor terms. In particular, by flipping the spins in rows $1$ and $2$ (Fig.~S2) after time $T_{d}/2$ where $T_{d}$ is a small fraction of the desired evolution duration, the next-nearest-neighbor interactions are refocused since each term contains spins from only row $1$ or $2$. However, half of the nearest neighbor interactions are also refocused, leaving effective evolution under the Hamiltonian $H_{eff1} = \sum \kappa_{N} ( S_{A}^{+}S_{B}^{-} +S_{C}^{+}S_{D}^{-}) + \text{h.c.}$ Analogously, by flipping the spins in rows $2$ and $3$, effective evolution under the Hamiltonian $H_{eff2} = \sum \kappa_{N} ( S_{B}^{+}S_{C}^{-} +S_{D}^{+}S_{A'}^{-}) + \text{h.c.}$ is achieved, again with $H_{NN}$ refocused. Combining the evolution according to $H_{eff1}$ and $H_{eff2}$ yields the desired nearest-neighbor Hamiltonian with next-to-nearest neighbor interactions refocused. However, since $H_{eff1}$ and $H_{eff2}$ do not commute, it will be necessary to employ piecewise evolution according to the Trotter-Suzuki formalism~\cite{Trotter59X, Suzuki76X, Hatano05X}. Further refocusing of higher order non-nearest neighbor interactions can also be achieved by extending the number of rows corresponding to the saw-tooth design; such an extension allows for the isolation of each specific pair of nearest neighbor interactions, thereby achieving the desired nearest-neighbor evolution through a Trotter sequence. 

\subsubsection*{Control of Directionality in Free Fermion State Transfer}

\noindent We consider the Hamiltonian presented in Eq.~{\bf{4}} in the main text. Under the assumption that $g \ll \frac{\kappa}{\sqrt{N}}$, we work perturbatively in eigenstates of the bare Hamiltonian $H_{0} = \sum_{i=1}^{N-1} \kappa (S_{i}^{+} S_{i+1}^{-} + S_{i}^{-} S_{i+1}^{+}) $, and consider coupling through the perturbation Hamiltonian $H' = g(S_{NV_{1}}^{+} S_{1}^{-} + S_{NV_{2}}^{+} S_{N}^{-} + \mbox{ h.c. })$. The essence of FFST can be understood as the long-range coherent interaction between the spin qubits, mediated by a specific collective eigenmode of the intermediate spin chain. This mode is best understood via Jordan-Wigner (JW) fermionization~\cite{Jordan28X, Bethe31X, Lieb61X}, which allows for the states of an XX spin chain to be mapped into the states of a set of non-interacting spinless fermions. In this representation, the state transfer is achieved by free fermion tunneling. By ensuring that the end spin qubits are weakly coupled to the intermediate spin chain, it is possible to tune the NV registers to achieve resonant tunneling through only a single particular fermionic eigenmode~\cite{YJG10X}.  Particle-hole symmetry of the Hamiltonian implies that the energy spectrum of the single fermion manifold is mirror symmetric across $E=0$. This implies that in the case of even $N$ intermediate spin chains (which are uniquely present in the proposed architecture), the NV registers are always initially off-resonant from all fermionic eigenmodes of both left and right spin chains. 

In this context, directionality becomes easily achievable so long as the left and right spin chains are of differing lengths, since the single fermion spectra will then be distinct. By ensuring that $g$ is sufficiently small, it is possible to tune to and hence be resonant with only a single spin chain direction. In particular, the energy spectrum of the single fermion manifold is given by $E_{k} = 2 \kappa \cos \frac{k \pi}{N+1}$, where $k=1, \cdots, N$; thus, tuning to a particular fermionic eigenmode corresponds to making choices of $\Delta$ and $\Omega$ which ensure that $\Delta  - \Omega_N \approx E_{k}$. The specific choice of $k$ corresponds to the particular single fermion eigenmode which is being tuned to. Interestingly, this can allow for control over the speed of quantum state transfer~\cite{YJG10X}. This speed is maximimized for $k=\frac{N}{2}\pm1$ in the case of even $N$ chains and for $k=\frac{N+1}{2}$ in odd $N$ chains. Directional control over state transfer is achieved by ensuring that only either the left or right DSCB is resonant with the NV register.  In this scenario, the coupling between the NV and the neighboring spin chain, which is off-resonant is highly suppressed. Assuming that the two neighboring spin chains are of differing lengths $N_{1}$ and $N_{2}$, the characteristic energy separation between fermionic eigenmodes in the two chains is approximately $\frac{\kappa}{N_{1}} - \frac{\kappa}{N_{2}} = \frac{\kappa (N_{1}-N_{2})}{N_{1} N_{2}}$. Thus, by ensuring that the register-impurity coupling $g$ is smaller than such an energy separation, it is possible to ensure that only single directional FFST occurs. Additionally, such an analysis suggests that by tuning $g$, it may be possible to overcome coupling-strength disorder induced by imperfect impurity implantation~\cite{YJG10X}. In particular disorder will cause localization, asymmetry of the eigenmodes, and changes in the statistics of the eigenenergies.  In the case of coupling-strength disorder, there exists an extended critical state at $E=0$ with a diverging localization length; this ensures the existence of an extended eigenmode with a known eigenenergy, suggesting that FFST is intrinsically robust against coupling-strength disorder~\cite{Evers08X, Balents97X}. However, the existence of an extended mode is not sufficient to ensure state transfer as disorder also enhances off-resonant tunneling rates and causes the eigenmode wavefunction amplitude to become asymmetric at the two ends of the chain. Despite such imperfections, by individually tuning the qubit-chain couplings, $g_{left}$ and $g_{right}$, it is possible to compensate for eigenmode asymmetry; furthermore, sufficiently decreasing the magnitude of the qubit-chain coupling ensures that off-resonant tunneling can safely be neglected, even in the presence of disorder~\cite{YJG10X}.  

\subsubsection*{Full Frequency Spectrum in Super-plaquettes} 

\noindent As mentioned in the main text, the frequency spectrum of a given super-plaquette requires careful consideration. In particular, given a sufficiently large magnetic field gradient, it will be essential to consider the full range of frequencies covered throughout a super-plaquette. For example, although the NV zero-field splitting (ZFS) $\Delta_{0} \sim 3$GHz naively protects the register from off-resonant excitations during impurity manipulation, it is important to note that this static ZFS is bridged as rows progress vertically along the field gradient. Given a gradient of order $100$MHz$/10$nm (in this section, for simplicity, we assume $10$nm row separation), a displacement of only 30 rows along the gradient direction is sufficient to fully bridge the NV ZFS. Furthermore, since each Nitrogen impurity requires four separate frequencies to be addressed, it is essential to ensure that such a congested frequency spectrum within a super-plaquette does not render the error probability intolerable. As discussed, the four frequencies corresponding to each Nitrogen impurity result from the two possible nuclear spin states and the choice of whether the JT axis lies along the NV axis; hence, the possible frequencies are $\omega_{0} \pm 1/2   A_{\parallel} =\omega_{0} \pm 80$MHz and $\omega_{0}\pm 1/2 (\frac{A_{\parallel}}{9} + \frac{8  A_{\perp}}{9}) = \omega_{0}\pm 60$MHz. To describe an adequate framework regarding the frequency spectrum, it is possible to consider a simplified problem, in which we ignore the nuclear spin and thus, are left with three transitions corresponding to the two NV transitions ($|0\rangle \rightarrow |\pm1\rangle$) and the Nitrogen impurity transition. For a static magnetic field that gives a Nitrogen ESR frequency equal to $\omega_0$, the NV transitions correspond to $\omega_0 + 3$GHz and $\omega_0 - 3$GHz. Thus, we can describe the situation according to three base lines at $-3,0,3$GHz, where the full frequency spectrum within the super-plaquette results from the addition of all possible lines at subsequent $10$nm intervals. 

One possible solution to avoid frequency overlap is to choose the field gradient to be $3*\zeta$MHz$/10$nm, where $\zeta$ is as yet undetermined. The possible choices of $\zeta$ correspond to field gradients which for some $n \in \mathbb{Z}$ yield that after $n*10$nm, the field due to the gradient is $3$GHz$-\zeta$. Such a choice immediately ensures that (disregarding JT) all possible transitions within a super-plaquette are non-overlapping. Furthermore, by correctly choosing $\zeta$, it is possible to ensure that even within the context of addressing the four individual Nitrogen impurity transitions, the separation of any pair of nearest frequencies is $\sim 10$MHz. A simple example can be illustrated by considering a gradient with strength $150$MHz$/10$nm, where the absolute minimum spacing between any two frequencies in a super-plaquette is $\Delta_{g} = 10$MHz. 

\vspace{5mm}
\subsection*{Proposed Experimental Realization and Limitations}

\subsubsection*{NV Implantation and Conversion} 

\noindent Although we have considered various errors accumulated during single- and two-qubit gate operations, an important further consideration is the error associated with the imperfect positioning of Nitrogen ion implantations. This imperfect implantation leads to errors in both the coupling strength as well as the individual impurity ESR frequencies. Crucially, both the vertical adiabatic sequential SWAP and the horizontal FFST are robust to fluctuations in coupling strength. The adiabatic sequential SWAP is also robust to errors induced by ESR frequency variations. Such variations result in an effective Rabi frequency, which alters the start and end point of the optimized ramp profile, without significantly affecting the adiabatic passage~\cite{Roland02X}. On the other hand, such ESR frequency fluctuations induce an off-resonant error during FFST; however, improvements in the implantation precision and utilizing larger $\Omega_{N}$ can sufficiently suppress such errors. 

\vspace{5mm}

\noindent In addition to implantation errors, it is also important to consider the infidelity induced by imperfect NV conversion efficiency. In particular, current experiments are limited by an optimistic NV conversion efficiency of approximately $40\%$. Within the context of the current architecture design (Fig.~S2), each plaquette contains 8 possible implantation sites that will allow for a staggered super-plaquette NV register structure, yielding a nominal error below $10^{-2}$. Thus, each super-plaquette will have a filling fraction greater than $99\%$ of functional plaquettes and the gate overhead associated with such faulty qubits will be negligible~\cite{Fowler10X}. Since the threshold penalty for faulty qubits is expected to be proportional to the gate overhead, the error threshold $\epsilon$ is expected to remain $\approx 1.4\%$~\cite{Fowler10X,Wang10bX}.  

\vspace{5mm}

\noindent Furthermore, the zig-zag structure of the vertical impurity spin chain allows for connectivity between any NV lattice site and the lattice site directly above in the corresponding vertically adjacent plaquette, \emph{as well as} any of the four nearest neighbor lattice sites in either direction (in order to account for imperfect conversion efficiency), as can be seen in Fig.~S2. The implantation region of vertically adjacent plaquettes is somewhat limited by the particular location of NV registers in the preceding vertical plaquette; however, such errors can be made negligible by considering larger plaquette sizes and hence, a larger range of connected vertical sites through the impurity spin chain.

\end{document}